\documentclass{Rinton-P9x6}

\newcommand\newblock{\hskip .11em}

\begin{document}

\title{Quantum information science and complex quantum systems}

\author{Michael A. Nielsen}

\address{Department of Physics \\ The University of Queensland,
Queensland 4072, Australia
\\
E-mail: nielsen@physics.uq.edu.au}


\maketitle

\abstracts{ What makes quantum information science a \emph{science}?
  This paper explores the idea that quantum information science may
  offer a powerful approach to the study of \emph{complex quantum
    systems}.}

\section{Introduction}

My subject in this paper is using quantum information science as an
approach to the study of complex quantum systems.  The work I describe
has involved many collaborators at the University of Queensland and
MIT, but I would especially like to emphasize the contribution of
Tobias Osborne.

Let me begin by asking what it is that makes quantum information
science a science?  Friends outside the field sometimes comment that
it seems to be largely engineering, with little science.  A standard
response from physicists is that in the course of building devices
like quantum computers, we'll discover lots of interesting physics.
This is undoubtedly true, and is an excellent reason for doing quantum
information science.  But it seems a little like the argument
sometimes used to justify going to the moon, namely, that it resulted
in valuable spin-off technologies in fields such as computation and
aeronautical engineering.  This misses a large part of the point,
since going to the moon has an \emph{intrinsic worth}, a point obvious
even to a small child.

What is the intrinsic worth of quantum information science?  In this
paper I argue, that quantum information science is a powerful approach
to the study of \emph{complex quantum systems}.  Related ideas have
been advocated previously by many people.
Aharonov~\cite{Aharonov99b}, Nielsen~\cite{Nielsen98d} and
Preskill~\cite{Preskill00a} argued that there may be connections
between the quantitative theory of entanglement and many-body quantum
systems, and there is a burgeoning literature exploring these
connections.  More explicitly, the concluding paragraph of
DiVincenzo's~\cite{Divincenzo00a} paper on the physical requirements
for quantum computation suggests that quantum information science may
offer valuable insights into complex quantum systems.  This theme was
explored in more detail by Osborne and
Nielsen~\cite{Nielsen01d,Osborne02b,Osborne02a,Osborne02c}, and the
present paper is an outgrowth of this work.

\section{Complex quantum systems}

What is a complex system?  Complexity is an elusive concept: it is
difficult to define, but we know it when we see it.  In response to
this difficulty one might ask whether it is possible to quantify
complexity.  In the 1980s Bennett proposed a measure of complexity
called the \emph{logical depth}~\cite{Bennett90a}.  The idea is that a
system is complex, or logically deep, if a description of the system
can be generated by a few simple rules, but those rules require a long
time to run.  For example, a human body is complex because it is
specified by a relatively small amount of information encoded in DNA,
but it takes a great deal of processing to get from DNA to the human
body.  Another example is a regular pattern on a checkerboard, which
is not complex because it can be quickly generated by a simple rule.
More subtle is the case of a random pattern on a checkerboard.  That
is not complex either, because there is no simple rule generating the
pattern.  Indeed, the simplest rule generating the pattern is simply
the program which contains (and prints) a complete listing of the
states of all the elements of the checkerboard, and this program runs
very quickly.

\begin{figure}[t]
\epsfig{file=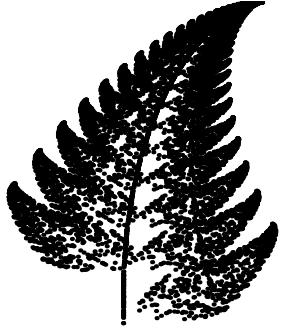,height=1.7in}
\caption{The fern is formed by iterating the rule 
in Eqs.~(\ref{eq:fern1})-(\ref{eq:fern4}) a few thousand times.
\label{fig:fern}}
\end{figure}

Let me give an example of something complex, that is, with high
logical depth.  Suppose we take a point, $x$, in the plane, for example,
$x=(0,0)$.  We bounce the point around the plane by repeatedly
applying one of the following four rules at random~\cite{Barnsley88a},
\begin{eqnarray}
 x \rightarrow \left[ \begin{array}{cc} 
    0 & 0 \\ 0 & 0.16 \end{array} \right]x & &
\mbox{with probability } 0.01  \label{eq:fern1} \\
 x \rightarrow \left[ \begin{array}{cc}  
    0.85 & 0.04 \\-0.04 & 0.85 \end{array} \right]x 
    + \left[ \begin{array}{c} 0 \\ 1.6 \end{array} \right]
    & & \mbox{with probability } 0.85 \\
 x \rightarrow \left[ \begin{array}{cc} 
    0.2 & -0.26 \\ 0.23 & 0.22 \end{array} \right]x 
    + \left[ \begin{array}{c} 0 \\ 1.6 \end{array} \right]
    & & \mbox{with probability } 0.07 \\
  x \rightarrow \left[ \begin{array}{cc} 
    -0.15 & 0.28 \\ 0.26 & 0.24 \end{array} \right]x 
    + \left[ \begin{array}{c} 0 \\ 0.44 \end{array} \right]
    & & \mbox{with probability } 0.07. \label{eq:fern4}
\end{eqnarray}
When this procedure is repeated a few thousand times an interesting
thing happens: with very high probability a fern shape fills in, as
illustrated in Fig.~\ref{fig:fern}. Furthermore, the fern is complex
in that there is a simple rule generating the fern, but it takes a
long time to run.

Bennett formalized these intuitions by defining the logical depth of a
data set to be the running time of a near-optimal computer program
generating that data set.  There are some technicalities hidden in
this definition, like the precise meaning of ``near-optimal'', that I
will gloss over.  Nonetheless, I would like to give the intuitive
flavour of the definition.  By ``near-optimal'' we mean that the
program is nearly the shortest possible.  The motivating idea is an
analogy between computer programs generating data sets and scientific
hypotheses.  Scientists tend to prefer simple explanations over more
complex, so if we think of computer programs as explanations for data
sets, then we would prefer short computer programs --- simple
``explanations'' --- over longer programs.  With this definition,
simple repeated structures and random patterns, like the checkerboards
described earlier, have low logical depth.  Systems like the fern have
high logical depth because they have simple explanations that take a
long time to run.

There is an interesting quantum twist to logical depth.  As we know,
factoring integers is a hard problem, hard enough that RSA systems
offers lots of money to people able to factor large integers: US
\$200,000 for a 2048-bit integer, at the time of writing.  Let's
optimistically imagine that it's ten years from now and somebody wants
to prove that they have a functioning quantum computer in their lab,
but don't want to reveal the details of how they built it.  One good
way of proving this would be to publish a paper containing the prime
factors of a large group of big integers --- perhaps the prime factors
of all numbers between $10^{1000}$ and $10^{1000}+1000000$.

Is this list of factors a complex system?  The answer depends on
whether the computer in the definition of logical depth is quantum or
classical.  If it's quantum then it seems likely that this system is
\emph{not} logically deep, and thus not complex, because we can
quickly generate the list using a short quantum program, namely,
Shor's algorithm~\cite{Shor97a}.  If the computer is classical, and
there really is no fast classical factoring algorithm, then the list
of factors has high logical depth, since there are simple computer
programs capable of generating such a list, but they operate very
slowly.

Thus, there are two distinct notions of logical depth, classical
logical depth, and quantum logical depth.  We can summarize the
situation by dividing systems into three distinct types.  First, there
are systems which have both low classical logical depth and low
quantum logical depth; these are ``simple''.  Then there are systems
that have high classical logical depth, but low quantum logical depth,
like the list of prime factors discussed above; these systems might be
called ``classically complex''.  Finally, there are systems with both
high classical logical depth and high quantum logical depth; the truly
``quantum complex'' systems.  I don't know of any examples with this
property, but consider some systems likely candidates, for example,
the output of a quantum cellular automata that's been running a long
time.  Note that systems with high quantum logical depth but low
classical logical depth seem unlikely, because a simple, fast
classical computer generating a data set can be simulated by a simple,
fast quantum computer.

\section{Quantum dynamics as a physical resource}

I've talked about quantifying complexity in quantum systems, and the
information-theoretic viewpoint has led us to the idea that there are
at least two, if not more, qualitatively different types of complex
system.  I'd like now to talk about what we can learn about specific
complex quantum systems from quantum information science.

Over the past few years great effort has been devoted to developing a
quantitative theory of quantum entanglement.  In my opinion a major
area in which this theory will be applied is to obtain insights into
the properties of complex quantum systems.  However, static quantum
entanglement is only a small part of the story: it is also interesting
to obtain a better understanding of the quantum \emph{dynamics} of
complex systems.  To achieve this this, my group has focused on
quantifying the \emph{strength} of a quantum dynamical operation for
information processing.

The motivation for this idea is the observation that quantum dynamics
are a \emph{fungible} physical resource, in the sense that it is
possible to interconvert different dynamical operations, just as it is
possible to convert one type of entangled state to another.  More
precisely, suppose a system contains $n$ qudits, and the Hamiltonian
for the system contains only two-body terms, and so can be represented
by a graph whose vertices represent qudits, and whose edges represent
the presence of an interaction between those qudits.  Finally, suppose
the graph is connected, so different qudits aren't cut off from one
another.  It turns out that by alternating evolution due to such a
Hamiltonian with single-qudit gates, we can, in principle, efficiently
simulate any quantum
computation\footnote{See~\cite{Dodd02a,Bennett01a,Wocjan02b}, and
  references therein.}.

While theoretically interesting, until recently most of these schemes
were not practically useful, requiring extremely frequent local
control to do simple operations such as the {\sc cnot}.  Even an
optimistic example~\cite{Dodd02a} required $\approx 10^4$ operations
to do a {\sc cnot}.

Recently, the situation has changed.  J.-L. and R.
Brylinski~\cite{Brylinski02a} have shown that, given any entangling
two-qudit unitary $U$, and local unitaries, it is always possible to
do universal quantum computation.  However, their proof used ideas
from algebraic geometry and the theory of Lie algebras, and it is not
clear to me whether their proof implies an efficient constructive
method for doing the {\sc cnot}.  Bremner \emph{et al}
\cite{Bremner02a} built on this work by giving a \emph{simple,
  constructive algorithm} for doing a {\sc cnot}, and thus universal
quantum computation, using any entangling two-qubit unitary operation,
and local unitaries.  The algorithm of \cite{Bremner02a} turns out to
be \emph{near-optimal}, using nearly the minimal possible number of
uses of $U$ to simulate a {\sc cnot}, and thus shows that quantum
dynamical operations are fungible not only in principle, but may be in
practice as well.

Knowing that quantum dynamics are a fungible physical resource, and
thus qualitatively equivalent to one another, we can try to quantify
the \emph{strength} of a dynamical operation.  We will now examine
some strength measures, basing our discussion on~\cite{Nielsen02a}.
Let's start by defining a strength measure for an $n$-qubit unitary
operation, $U$.  (This is just one example among many
in~\cite{Nielsen02a}.)  A metric, $D$, on unitary operations induces a
natural strength measure, $K_D(U) \equiv \min_{A,B,\ldots}
D(U,A\otimes B \otimes \ldots)$.  That is, the strength is the minimal
distance between $U$ and the set of local unitaries.

What good is a strength measure?  Let me answer by explaining a
connection between strength and computational complexity.  Imagine we
have a strength measure with the following three properties.  The
first property, \emph{chaining}, says that the strength of a product
of two unitary operations, $U$ and $V$, is less than the sum of their
combined strengths, $K(UV) \leq K(U)+K(V)$.  The intuition is that the
ability to do $U$ and $V$ separately should be \emph{at least} as
powerful as the ability to do $UV$.  This property is not always true
for the metric-based strength measures, but is true for a large
subclass\cite{Nielsen02a}.

The second property, \emph{stability}, says that if we add an extra
qubit to our system and do nothing to it, that should not change the
strength, $K(U \otimes I) = K(U)$.  The metric-based measures do not
always satisfy stability, but they do in some instances.  The third
property, \emph{locality}, says that a strength measure should be zero
for products of local unitary operations, $K(A \otimes B \otimes
\ldots) = 0$.  This is true for the metric-based measures of strength.

Imagine $K$ is a strength measure satisfying these properties, and we
want to perform a unitary, $U$, using {\sc cnot} and single-qubit
gates.  Imagine the circuit contains $M$ {\sc cnot}s.  Applying the
properties, $K(U)$ can be no more than the sum of the strengths of the
{\sc cnot}s, so $M \geq K(U)/K(\mbox{{\sc cnot}})$.  By stability, the
strength of {\sc cnot} is constant, so if $K(U)$ scales
superpolynomially, then so must the number of gates needed to do $U$.
Thus, developing measures of dynamic strength may enable progress on
the notoriously difficult problem of proving lower bounds on
computational complexity, and thus to gain insight into the enormously
complex space of quantum dynamical processes.

Let me conclude by returning to the big picture.  I believe that the
major scientific task of quantum information science is to develop
tools for the study of complex quantum systems.  In quantum mechanics
we're like chess players who've learnt the rules of the game, but are
still trying to figure out the emergent properties those rules imply.
We're doing so by developing overarching theories, like the theory of
entanglement and of dynamic strength, which let us understand ever
more complex phenomena.  I expect that as these theories are developed
they will enable us to better understand complex systems, not only in
information processing, but also in other areas of many-body physics.

\section*{Acknowledgements}

The point of view in this paper owes a lot to many enjoyable
conversations with Jennifer Dodd and Tobias Osborne.  Thanks also to
Andrew Childs and Tobias Osborne for feedback on an earlier version of
this manuscript~\cite{Nielsen02b}.


\begin{thebibliography}{10}

\bibitem{Aharonov99b}
D.~Aharonov.
\newblock {\em Phys. Rev. A}, 62(6):2311, 1999.

\bibitem{Nielsen98d}
M.~A. Nielsen.
\newblock {\em Quantum Information Theory}.
\newblock PhD thesis, University of New Mexico, 1998.
\newblock {arXiv}:quant-ph/0011036.

\bibitem{Preskill00a}
J.~Preskill.
\newblock {\em J. Mod. Opt.}, 47(2):127--137, 2000.
\newblock {arXiv}:quant-ph/9904022.

\bibitem{Divincenzo00a}
D.~P. DiVincenzo.
\newblock {\em Fortschritte der physik --- progress of physics},
  48(9--11):771--783, 2000.
\newblock {arXiv}:quant-ph/0002077.

\bibitem{Nielsen01d}
M.~A. Nielsen.
\newblock Seminar at the KITP, UC Santa Barbara (2001).
\newblock http://online.kitp.ucsb.edu/online/qinfo01/nielsen/.

\bibitem{Osborne02b}
T.~J. Osborne.
\newblock {\em Entanglement and complex quantum systems}.
\newblock PhD thesis, submitted at the University of Queensland, 2002.

\bibitem{Osborne02a}
T.~J. Osborne and M.~A. Nielsen.
\newblock {arXiv}:quant-ph/0202162, 2002.

\bibitem{Osborne02c}
T.~J. Osborne and M.~A. Nielsen.
\newblock {\em Quantum Information Processing}, 1(1):45--53, 2001.
\newblock {arXiv}:quant-ph/0109024.

\bibitem{Bremner02a}
M.~J. Bremner, C.~M. Dawson, J.~L. Dodd, A.~Gilchrist, A.~W. Harrow,
  D.~Mortimer, M.~A. Nielsen, and T.~J. Osborne.
\newblock {\em {arXiv}:quant-ph/0207072}, 2002.

\bibitem{Nielsen02a}
M.~A. Nielsen, C.~M. Dawson, J.~L. Dodd, A.~Gilchrist, D.~Mortimer, T.~J.
  Osborne, M.~J. Bremner, A.~W. Harrow, and A.~Hines.
\newblock {\em {arXiv}:quant-ph/0208077}, 2002.

\bibitem{Bennett90a}
C.~H. Bennett.
\newblock In W.~H. Zurek, editor, {\em Complexity, entropy and the physics of
  information}, pages 137--148. Addison-Wesley, Redwood City, CA, 1990.

\bibitem{Barnsley88a}
M.~F. Barnsley.
\newblock {\em Fractals everywhere}.
\newblock Academic Press, 1988.

\bibitem{Shor97a}
P.~W. Shor.
\newblock {\em SIAM J. Comp.}, 26(5):1484--1509, 1997.

\bibitem{Bennett01a}
C.~H. Bennett, J.~I. Cirac, M.~S. Leifer, D.~W. Leung, N.~Linden, S.~Popescu,
  and G.~Vidal.
\newblock {\em {arXiv}:quant-ph/0107035v1}, 2001.

\bibitem{Wocjan02b}
P.~Wocjan, M.~Roetteler, D.~Janzing, and {Th.} Beth.
\newblock {\em Quantum Information and Computation}, 2(2):133--150, 2002.
\newblock {arXiv}:quant-ph/0109063.

\bibitem{Dodd02a}
J.~L. Dodd, M.~A. Nielsen, M.~J. Bremner, and R.~T. Thew.
\newblock {\em Phys. Rev. A}, 65(4):040301 (R), 2002.
\newblock {arXiv}:quant-ph/0106064.

\bibitem{Brylinski02a}
J.~L. Brylinski and R.~Brylinski.
\newblock {\em Universal quantum gates}, chapter II in~\cite{Brylinski02b}.
\newblock 2002.
\newblock ar{X}iv:quant-ph/0108062.

\bibitem{Brylinski02b}
R.~K. Brylinski and G.~Chen, editors.
\newblock {\em Mathematics of {Q}uantum {C}omputation}.
\newblock Computational Mathematics. Chapman \& Hall / CRC Press, 2002.

\bibitem{Nielsen02b}
M.~A.~Nielsen.
\newblock ar{X}iv:quant-ph/0208078.

\end{thebibliography}

\end{document}